\documentclass[prl,aps,superscriptaddress,twocolumn,amssymb,showpacs]{revtex4}
\usepackage{graphicx}
\usepackage{longtable}
\usepackage{epsfig}
\usepackage{dcolumn}
\usepackage{bm}

\begin{document}

\title{       Mechanism of the Verwey transition in magnetite }

\author{      Przemys\l{}aw Piekarz }
\affiliation{ Institute of Nuclear Physics, Polish Academy of Sciences, 
              Radzikowskiego 152, PL-31342 Krak\'{o}w, Poland }

\author{      Krzysztof Parlinski }
\affiliation{ Institute of Nuclear Physics, Polish Academy of Sciences, 
              Radzikowskiego 152, PL-31342 Krak\'{o}w, Poland }
	  
\author{      Andrzej M. Ole\'{s} }
\affiliation{ Institute of Nuclear Physics, Polish Academy of Sciences, 
              Radzikowskiego 152, PL-31342 Krak\'{o}w, Poland }
\affiliation{ Max-Planck-Institut f\"ur Festk\"orperforschung,
              Heisenbergstrasse 1, D-70569 Stuttgart, Germany }

\begin{abstract}
By combining {\it ab initio\/} results for the electronic structure 
and phonon spectrum with the group theory, we establish the origin 
of the Verwey transition in Fe$_3$O$_4$. Two primary order parameters
with $X_3$ and $\Delta_5$ symmetries are identified. They induce the 
phase transformation from the high-temperature cubic to the 
low-temperature monoclinic structure. The on-site Coulomb interaction 
$U$ between $3d$ electrons at Fe ions plays a crucial role in this 
transition --- it amplifies the coupling of phonons to conduction 
electrons and thus opens a gap at the Fermi energy. \\
{\it Published in Phys. Rev. Lett. {\bf 97}, 156402 (2006).}
\end{abstract}

\date{\today}
\maketitle 

The discovery of the remarkable discontinuous drop of the electrical 
conductivity by two orders of magnitude in magnetite (Fe$_3$O$_4$) below 
the Verwey transition (VT) at $T_V=122$ K \cite{verwey} triggered 
intensive studies of its microscopic origin which have been continued 
for more than six decades. Verwey suggested explaining it by a 
metal-insulator transition due to the reduction of electron mobility 
caused by ordering of Fe$^{3+}$ and Fe$^{2+}$ ions below $T_V$. In spite 
of great experimental effort, however, the existence of ionic-like 
charge ordering at $T<T_V$ could not be confirmed, and the origin of 
the VT in magnetite remains still puzzling \cite{review}. 

Magnetite is a ferrimagnetic spinel with anomalously high critical 
temperature $T_c\simeq 860$ K. Hence, it is viewed as an ideal candidate 
for room temperature spintronic applications. It crystalizes in the 
inverse spinel cubic structure, Fd${\bar 3}$m, with two types of Fe 
sites: the tetrahedral $A$ sites and the octahedral $B$ ones, see Ref. 
\cite{Wri02}. Below $T_c$, Fe$_3$O$_4$ is magnetically ordered with 
antiparallel moments at $A$ and $B$ sites. Within the commonly used 
ionic picture $A$ sites are occupied by Fe$^{3+}$ ions, and $B$ sites 
are occupied randomly by Fe$^{3+}$ and Fe$^{2+}$ ions at room 
temperature. The Verwey model assumes a purely electronic mechanism of 
the VT, leading below $T_V$ to ordering of Fe$^{2+}$ and Fe$^{3+}$ ions 
in $B$-chains, along [110] and [1$\bar{1}$0] directions, respectively. 
Anderson \cite{anderson} argued that the charge ordering in a form of a 
charge density wave, is stabilized by interionic electrostatic energy, 
where each $B$-tetrahedron consists of two Fe$^{2+}$ and two Fe$^{3+}$ 
ions. This picture was studied in the Hubbard-like model with 
interatomic $d-d$ Coulomb interactions \cite{cul-cal}, but was not confirmed 
by nuclear magnetic resonance and M\"{o}ssbauer measurements \cite{garcia}. 
While recent resonant soft X-ray scattering suggests charge-orbital ordering within 
the oxygen $2p$ states \cite{Hua06} and a cooperative Jahn-Teller 
effect \cite{Sch05}, the microscopic mechanism of the VT remains controversial.

There are also other arguments against simple ionic mechanism of the VT. 
Firstly, the replacement of oxygen O$^{16}$ by O$^{18}$ isotope 
increases $T_V$ by a few degrees \cite{isotop}, indicating that the 
transition cannot be purely electronic. Secondly, signals in diffuse 
neutron scattering observed above $T_V$ at 
${\bf k}_{\Delta}=(0,0,\frac{1}{2})$, halfway between $\Gamma $ and $X$ 
points with ${\bf k}_X=(0,0,1)$, and equivalent points of the reciprocal 
lattice \cite{diff1} (in units of $\frac{2\pi}{a}$, where $a$ is the 
cubic lattice constant), change into Bragg peaks below $T_V$. The 
intensity of critical scattering was succesfully calculated on the basis 
of the transverse acoustic (TA) phonon mode with the $\Delta_5$ 
symmetry. Further neutron studies discovered diffuse scattering with 
large intensities close to $\Gamma$ and $X$ points, and the dominant 
component of $X_3$ symmetry \cite{diff3}. Other observations from Raman 
\cite{raman}, X-ray \cite{exafs} and nuclear inelastic scattering 
measurements \cite{handke} suggest that phonons play an essential role 
in the VT. But phonons alone cannot explain the metal-insulator nature 
of the transition either. Actually, if they alone were responsible, a 
phonon soft mode would have been found for the cubic phase, while the 
present {\it ab initio\/} calculations do not imply such a soft mode 
behavior.

The purpose of this Letter is to resolve the above controversies and to
demonstrate a cooperative scenario of the VT. Following the pioneering 
work of Ihle and Lorenz \cite{ihle}, who considered weak intersite 
Coulomb interactions \cite{Ima98}, we argue that a combination of 
local Coulomb interactions between $3d$ electrons and the electron-phonon 
(EP) coupling is the key feature responsible for the observed VT. 
Before presenting the evidence we recall that the monoclinic structure 
at low temperature \cite{iizumi}, space group P2/c, with unit cell close 
to $a/\sqrt2\times a/\sqrt{2}\times 2a$, was observed using the high 
resolution neutron and synchrotron X-ray diffraction 
\cite{Wri02}. 
Having P2/c symmetry of the low temperature phase, we get a consistent 
description of the VT, with reflections arising below $T_V$ at 
reciprocal lattice point ${\bf k}_{\Delta}$. Reflections at ${\bf k}_X$ 
and equivalent points are related to the unit cell twice shorter, 
namely $a/\sqrt2\times a/\sqrt{2}\times a$. Under typical circumstances 
one would select $\Delta_5$ symmetry as a primary order parameter (OP), 
while $X_3$ as a secondary one. The symmetry analysis, however, 
prohibits the $X_3$ symmetry to be a secondary OP of $\Delta_5$.

The electronic structures of magnetite in the cubic and monoclinic 
magnetically ordered phases \cite{Wri02} have been obtained using 
{\it ab initio\/} methods. In the cubic phase, the ground state is 
metallic with the 
minority spin $t_{2g}$ Fe($B$) states at the Fermi energy \cite{band}. 
The calculations for the monoclinic P2/c structure stable below $T_V$, 
performed using the LDA+$U$ method \cite{Leo04}, revealed the insulating 
state with a small gap of $0.18$ eV, the value being remarkably close to 
the experimental one $0.15$ eV \cite{gap_exp}. The gap opening is 
accompanied by rather subtle charge-orbital ordering, with the charge 
modulation amplitude in $3d$ Fe($B$) states of order $0.1e$, exactly the 
same as the one derived from diffraction analysis \cite{Wri02}, 
and below the sensibility of other experimental techniques \cite{garcia}.

\begin{table}[b!]
\caption{
List of OPs (IR) from the parent space group Fd${\bar 3}$m (No=227)
and basis $(1,0,0), (0,1,0), (0,0,1)$ 
to the monoclinic phase P2/c (No=13, unique axis $b$, choice 1),  
basis $(\frac{1}{2},-\frac{1}{2},0), (\frac{1}{2},\frac{1}{2},0),
(0,0,2)$ and origin $(\frac{1}{4},0,\frac{1}{4})$ relative to the 
original face center cubic lattice.
Size is the ratio of primitive low-symmetry to
high-symmetry unit cell volumes. Specific relationship between components
of the order parameters are not shown.}
\begin{ruledtabular}
\begin{tabular}{cccccc}
& IR & size & subgroup & No & \cr
\hline
& $\Gamma _1^+$ $(A_{1g})$ & 1  &  Fd$\bar 3$m & 227 & \cr
& $\Gamma _3^+$ $(E_{g})$ & 1   &  I$4_1$/amd  & 141 & \cr
& $\Gamma _4^+$ $(T_{1g})$ & 1  &  C2/m & 12 & \cr
& $\Gamma _5^+$ $(T_{2g})$ & 1  &  Imma & 74 & \cr
& $\Gamma _5^+$ $(T_{2g})$ & 1  &  C2/m & 12 & \cr
& $X_1$ & 2  &  Pmma & 51 & \cr
& $X_3$ & 2  &  Pmna & 53 & \cr
& $\Delta _2$ & 4 & Pcca & 54 & \cr
& $\Delta _4$ & 4 & Pcca & 54 & \cr
& $\Delta _5$ & 4 & Pbcm & 57 & \cr
\end{tabular}
\end{ruledtabular}
\label{table1}
\end{table}

In order to combine the observed and computed properties of magnetite to 
a consistent picture of the VT, we have performed a symmetry analysis 
based on the group theory. 
For this study, we used two computer codes: the COPL \cite{copl} 
and {\sc Isotropy} \cite{isotropy}. We recall that the high- and 
low-symmetry space groups for cubic and monoclinic phases 
\cite{Wri02} are Fd${\bar 3}$m and P2/c, respectively. The first 
observation is that there is {\it no} single irreducible representation 
(IR) of Fd${\bar 3}$m which lowers the space group Fd${\bar 3}$m 
directly to P2/c (see Tab. I). This means that the VT has to involve at 
least two OPs. Further symmetry analysis shows that two IRs, $\Delta_5$ 
and $X_3$, acting simulataneously, reduce the space group Fd${\bar 3}$m 
to P2/c. Indeed, there are two symmetry reduction relationships: 
Fd${\bar 3}$m $\rightarrow\Delta_5\rightarrow$ Pbcm(4) and
Fd${\bar 3}$m $\rightarrow$ $X_3$ $\rightarrow$ Pmna(2), 
where the increase of the unit cells is indicated in brackets. Common 
symmetry elements of Pbcm(4) and Pmna(2) form the space group P2/c(4). 
(We have verified that the symmetry elements are correctly located and 
oriented.) As a highly nontrivial result one finds that two independent 
IRs, $\Delta_5$ and $X_3$, give two primary OPs of the VT. Moreover, 
the $\Gamma^+_5=T_{2g}$ IR could also be involved in the VT. 
Its symmetry reduction relationship is 
Fd${\bar 3}$m $\rightarrow\Gamma^+_5\rightarrow$ Imma(1).
Symmetry elements of Imma(1) will not reduce further the symmetry of 
the space group P2/c(4). Thus, $\Gamma^+_5$ is a secondary OP. 
It represents a shear and contributes to the VT, as suggested by 
the observed softening of the $c_{44}$ elastic constant \cite{elastic}.

The group theory tells us the order of coupling terms between the OPs.
One finds that linear $\Delta_5\,\otimes X_3$, linear-bilinear 
$\Delta_5\,\otimes X_3\,\otimes X_3$ and 
$\Delta_5\,\otimes\Delta_5\,\otimes X_3$ terms are forbidden by 
symmetry. Therefore, the fourth order coupling 
$\Delta_5\,\otimes\Delta_5\,\otimes X_3\,\otimes X_3$ is the lowest one 
allowed by symmetry. The phase transition with two OPs has to be of the 
first order. 
The coupling between the secondary OP $\Gamma^+_5$ and $X_3$ is described
by the linear-bilinear $\Gamma^+_5\,\otimes X_3\,\otimes X_3$ term.
There are other IRs from $\Gamma $, $X$ and $\Delta$ 
reciprocal lattice points, which could become active at the VT as 
secondary OPs. Although they might further diminish the ground state 
energy, they cannot lower the crystal symmetry.

\begin{figure}[t!]
\includegraphics[width=7.7cm]{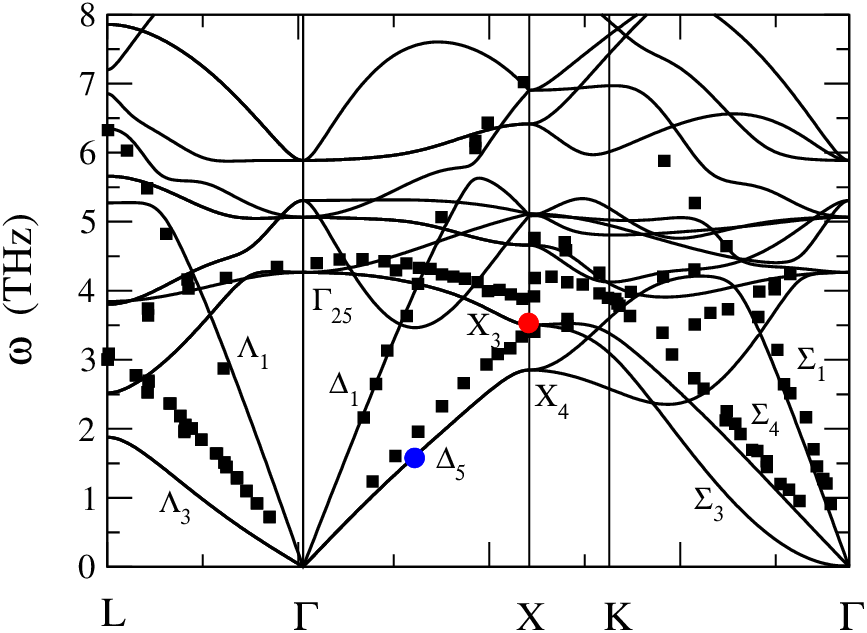}
\caption{(color online)
Low-frequency phonon dispersion relations for cubic phase of Fe$_3$O$_4$ 
calculated with $U=4$ eV and $J=0.8$ eV. The squares show the results of 
neutron scattering experiment \cite{phonons}. Two primary OPs are 
related to $\Delta_5$ and $X_3$ phonons marked by circles. The high symmetry 
points from left to right are: $L=(\frac{1}{2},\frac{1}{2},\frac{1}{2})$,
$\Gamma=(0,0,0)$, $X=(0,0,1)$, $K=(\frac{1}{4},\frac{1}{4},1)$, $\Gamma=(1,1,1)$}
\label{fig:fig1}
\end{figure}

The crystal structure of magnetite was optimized within the generalized 
gradient approximation (GGA) with on-site $3d$ electron interactions described 
by the Coulomb element $U$ and the Hund's exchange $J$ \cite{Ole83}, 
the so-called spin-polarized GGA+$U$ approach. The calculations were 
performed with the VASP code \cite{vasp} on two supercells: 
$a\times a\times a$ and approximately 
$a/\sqrt{2}\times a/\sqrt{2}\times 2a$, for the cubic and monoclinic 
structures, respectively, with 56 atoms each. We included six valence 
electrons for oxygen ($2s^22p^4$) and eight for iron ($3d^74s^1$) 
represented by plane waves with energy cut-off 520 eV. The wave 
functions in the core region were obtained by the full-potential 
projector augmented-wave method \cite{PAW}. The summation over the 
Brillouin zone was performed on the $4\times4\times4$ and 
$4\times4\times2$ ${\bf k}$-point grids for Fd${\bar 3}$m and P2/c 
symmetries, respectively. For the local interactions we choose the 
parameters \cite{Zaa90}: $U=4$ eV and $J=0.8$ eV.

Phonons were calculated only for the cubic phase, using the direct 
method implemented in {\sc Phonon} code \cite{direct}. The 
Hellmann-Feynman forces were obtained for six independent displacements: 
two for each nonequivalent atom in positive and negative directions  
(with the amplitude $0.02$ \AA). Using the respective force constants, 
the dynamical matrix was constructed and diagonalized. 
The selected $1\times1\times1$ supercell provides exact phonon
frequencies at the $\Gamma$ and $X$ points. The frequencies away from 
these points carry only negligible errors since the force constants 
decrease more than two orders of magnitude within the supercell.

The phonon dispersion curves, along the high-symmetry directions of the 
reciprocal space, are classified according to their IRs (Fig. 1). The 
longitudinal acoustic and TA modes in [001] direction correspond to 
one-dimensional $\Delta_1$ and two-dimensional $\Delta_5$ 
representations, respectively. As already mentioned, all phonon 
frequencies are real, and soft modes are absent in the cubic phase. 
The longitudinal acoustic modes are in a very good agreement with the 
neutron data measured at room temperature \cite{phonons}.
The discrepancies for the transverse phonons may result from the EP 
coupling, which effectively lowers phonon frequencies at low 
temperatures (as observed by diffuse scattering).

\begin{figure}[t!]
\includegraphics[width=7.5cm]{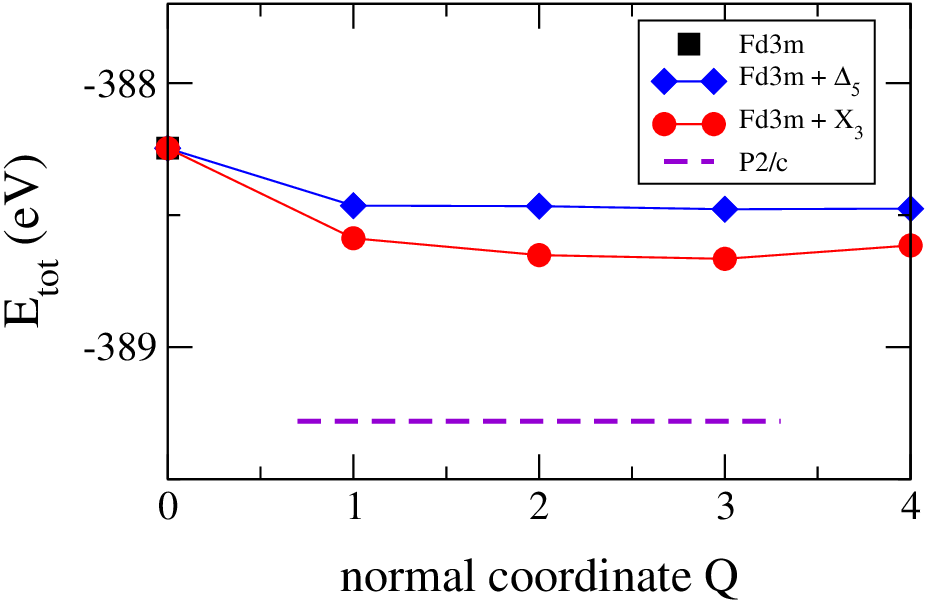}
\caption{(color online)
Total energies $E_{\rm tot}$ per supercell of 56 atoms for the 
structures: 
cubic Fd${\bar 3}$m, 
cubic with superimposed phonon modes of $\Delta_5$ at ${\bf k}_{\Delta}$ 
(Fd${\bar 3}$m+$\Delta_5$), and 
$X_3$ at ${\bf k}_X$ (Fd${\bar 3}$m+$X_3$) for increasing mode amplitude
$Q$, and for the ground state monoclinic P2/c. Parameters as in Fig. 1. 
}
\label{fig:fig2}
\end{figure}

The OPs $\Delta_5$ and $X_3$ could be (each one separately) a linear 
combination of the electronic and phononic components. In order to 
investigate the EP coupling for these modes, we computed the effect of 
applied lattice deformation, generated according to phonon polarization 
vectors, on the total energy and electron density of states. Among all 
the investigated modes with the $\Delta$ symmetry at ${\bf k}_{\Delta}$, 
only the TA ($\Delta_5$) mode shows a significant coupling to electrons 
(presented below). At the $X$ point, we found a strong EP interaction 
for the lowest transverse optic mode with the $X_3$ symmetry. 
In contrast, the TA ($X_4$) and higher optic phonons couple very 
weakly to electrons.

When the cubic crystal is distorted by either $X_3$ or $\Delta_5$ 
phonons, the total energy $E_{\rm tot}$ of the system decreases, 
indicating that these modes participate in the structural transition 
(Fig. 2). Note that if electrons were not involved, the ground state 
energies of the optimized structures distorted by phonons would have 
increased. The lowest energy was reached for the P2/c structure, 
with the optimized lattice constants and atomic positions close
to these of Ref. \cite{Wri02}, what confirms that monoclinic 
symmetry is stable at $T=0$ K. 

\begin{figure}[t!]
\includegraphics[width=7.5cm]{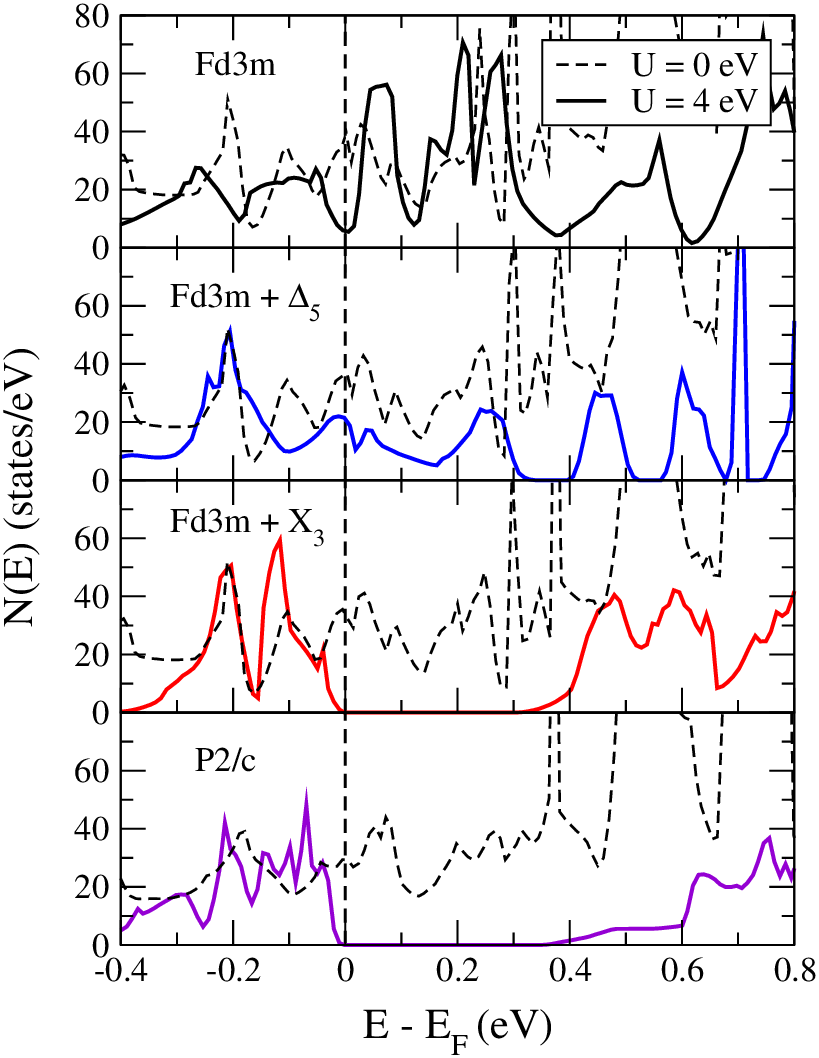}
\caption{(color online)
Electron minority density of states for the structures of Fig. 2: 
cubic Fd${\bar 3}$m, 
cubic with superimposed phonon modes of $\Delta_5$ at ${\bf k}_{\Delta}$ 
(Fd${\bar 3}$m+$\Delta_5$) and $X_3$ at ${\bf k}_X$ (Fd${\bar 3}$m+$X_3$) 
symmetry with normal coordinate $Q=3$, and monoclinic P2/c structures. 
Solid (dashed) lines correspond to $U=4$ eV and $J=0.8$ eV ($U=J=0$), 
respectively.
}
\label{fig:fig3}
\end{figure}

The crystal distortion which decreases the energy $E_{\rm tot}$ is 
directly connected with the opening of the gap at $E_F$ in the 
electronic density of states (Fig. 3). For two distorted structures, 
and for cubic Fd$\bar{3}$m and monoclinic P2/c ones, we compare the 
electronic structures obtained for $U=0$ and $U=4$ eV. In agreement with 
the previous calculations \cite{band}, only down spin $t_{2g}$ Fe($B$) 
states contribute at $E_F$, and the cubic phase is metallic 
independently of $U$. For the $\Delta_5$ deformation, magnetite remains 
a metal for $U=0$, while for $U=4$ eV one observes a significant 
reduction of the spectral weight above $E_F$, which does not yet lead 
to a gap. This indicates however a considerable enhancement of the EP 
coupling for the increasing Coulomb interaction $U$. In the case of the 
$X_3$ mode and for $U=4$ eV, the EP coupling is even much stronger ---
it opens a gap and triggers the metal-insulator transition. In contrast, 
a metallic phase is found for $U=0$. Finally, for the monoclinic P2/c 
symmetry, the electronic state is insulating (metallic) for $U=4$ eV 
($U=0$), in agreement with other calculations \cite{Leo04}. 

Altogether, the above results reveal a crucial role played by electron 
correlations in the VT. The strong Coulomb interaction $U$ reduces the 
mobility of electrons in $t_{2g}$ states, increasing their tendency 
towards localization. So modified electronic density responds to lattice 
deformation, leading to the electronic instability at the Fermi surface 
and to the gap opening. We emphasize that the present mechanism is novel 
and {\it does not} benefit from the Peierls-like distortions. Such a 
cooperative mechanism including strong electron correlations and the 
EP coupling was studied before \cite{fukuyama}, but the dimer-bond 
formation seems not to be supported experimentally \cite{Wri02}.

As argued by Wright {\it et al.\/} \cite{Wri02}, the lattice 
distortion and charge density following from diffraction analysis have 
predominantly character of $X$ modulation, with additional $\Delta$ 
modulation. This is in perfect agreement with 
our result, which shows a strong coupling of the $X_3$ phonon mode to 
the electronic density. In this mode, the Fe($B$) and O atoms are 
displaced along the diagonal [110] and [1$\bar{1}$0] directions, 
creating a polar deformation of $B$ site octahedra. This lattice 
distortion couples to charge fluctuations on Fe($B$) and O ions, 
inducing the diffuse scattering observed much above $T_V$. 
The observation that lattice deformation survives locally in the cubic 
phase \cite{exafs} indicates the existence of the precursor short-range 
order above $T_V$. It is further supported by photoemission spectroscopy, 
consistent with the reduction of the single particle gap to $\sim 0.1$ 
eV, not closing completely at $T>T_V$ \cite{gap_exp}. The long-range 
order, with larger gap, sets in at $T_V$ due to crystal-structural 
transformation, as was clearly demonstrated in recent high-pressure 
diffraction experiment \cite{high-pressure}. 
The observed charge disproportionation \cite{Hua06} results from
the modified $d$-$p$ hybridization due to the EP coupling.
Finally, a gap in the magnon spectrum at ${\bf k}_{\Delta}$ point below
$T_V$ \cite{McQ06} indicates a large spin-phonon interaction and confirms
that ${\bf k}_{\Delta}$ becomes a point on the Brillouin zone surface.

Summarizing, we have shown that the VT is driven by two primary OPs 
$\Delta_5$ and $X_3$, which are both characterized by strong linear EP 
coupling, amplified by Coulomb interactions. While the cubic phase 
would remain metallic without this coupling, charge fluctuations of two 
primary OPs induced by phonons simultaneously support local lattice 
distortions and open a pseudogap at the Fermi energy. As a result, 
the condensation of both OPs leads to the monoclinic distortion below 
the VT, a gap develops and the conductivity is lowered. At the 
structural transition, the charge modulation with a tiny amplitude is 
stabilized, being a manifestation of the common electronic and lattice 
origin of this transition.

{\it Note added.\/} After submitting this paper new resonant X-ray 
scattering data were published \cite{Naz06} which reveal fractional 
charge ordering below the VT.
They are fully compatible with the present explanation of the VT.

We acknowledge partial support by the European Community under FP6
contract No. NMP4-CT-2003-001516 (DYNASYNC).

\newpage

\end{document}